\documentclass[preprint2]{aastex}

\newcommand{\myemail}{enomoto@icrr.u-tokyo.ac.jp}

\slugcomment{To be submitted to ApjL.}

\shorttitle{Search for TeV gamma-rays from SN 1987A in 2001}
\shortauthors{Enomoto et al.}

\begin{document}

\title{Search for TeV gamma-rays from SN 1987A in 2001}

\author{
R.~Enomoto\altaffilmark{1},
L.T.~Ksenofontov\altaffilmark{1},
H.~Katagiri\altaffilmark{1},
K.~Tsuchiya\altaffilmark{1},
A.~Asahara\altaffilmark{2},
G.V.~Bicknell\altaffilmark{3},
R.W.~Clay\altaffilmark{4},
P.G.~Edwards\altaffilmark{5},
S.~Gunji\altaffilmark{6},
S.~Hara\altaffilmark{1,2},
T.~Hara\altaffilmark{7},
H.~Hattori\altaffilmark{8},
Sei.~Hayashi\altaffilmark{9},
Shin.~Hayashi\altaffilmark{9},
C.~Itoh\altaffilmark{10},
S.~Kabuki\altaffilmark{1},
F.~Kajino\altaffilmark{9},
A.~Kawachi\altaffilmark{1},
T.~Kifune\altaffilmark{11},
H.~Kubo\altaffilmark{2},
J.~Kushida\altaffilmark{2,12},
Y.~Matsubara\altaffilmark{13},
Y.~Mizumoto\altaffilmark{14},
M.~Mori\altaffilmark{1},
H.~Moro\altaffilmark{8},
H.~Muraishi\altaffilmark{15},
Y.~Muraki\altaffilmark{13},
T.~Naito\altaffilmark{7},
T.~Nakase\altaffilmark{8},
D.~Nishida\altaffilmark{2},
K.~Nishijima\altaffilmark{8},
M.~Ohishi\altaffilmark{1},
K.~Okumura\altaffilmark{1},
J.R.~Patterson\altaffilmark{4},
R.J.~Protheroe\altaffilmark{4},
K.~Sakurazawa\altaffilmark{12},
D.L.~Swaby\altaffilmark{4},
T.~Tanimori\altaffilmark{2},
F.~Tokanai\altaffilmark{6},
H.~Tsunoo\altaffilmark{1},
T.~Uchida\altaffilmark{1},
A.~Watanabe\altaffilmark{6},
S.~Watanabe\altaffilmark{2},
S.~Yanagita\altaffilmark{10},
T.~Yoshida\altaffilmark{10},
and T.~Yoshikoshi\altaffilmark{16}
}

\altaffiltext{1}{Institute for Cosmic Ray Research, University of Tokyo,
Chiba 277-8582, Japan}
\altaffiltext{2}{Department of Physics, Kyoto University, Kyoto 606-8502, Japan}
\altaffiltext{3}{MSSSO, Australian National University, ACT 2611, Australia}
\altaffiltext{4}{Department of Physics and Math. Physics, University of
Adelaide, SA 5005, Australia}
\altaffiltext{5}{Institute of Space and Astronautical Science, Kanagawa
229-8510, Japan}

\altaffiltext{6}{Department of Physics, Yamagata University, Yamagata 990-8560, Japan}
\altaffiltext{7}{Faculty of Management Information, Yamanashi Gakuin
University, Yamanashi 400-8575, Japan}
\altaffiltext{8}{Department of Physics, Tokai University, Kanagawa 259-1292, Japan}
\altaffiltext{9}{Department of Physics, Konan University, Hyogo 658-8501, Japan}
\altaffiltext{10}{Faculty of Science, Ibaraki University, Ibaraki 310-8512, Japan}
\altaffiltext{11}{Faculty of Engineering, Shinshu University, Nagano 380-8553, Japan}
\altaffiltext{12}{Department of Physics, Tokyo Institute of Technology,
Tokyo 152-8551, Japan}
\altaffiltext{13}{STE Laboratory, Nagoya University, Aichi 464-8601, Japan}
\altaffiltext{14}{National Astronomical Observatory of Japan, Tokyo 181-8588, Japan}
\altaffiltext{15}{Ibaraki Prefectural University of Health Sciences, Ibaraki 300-0394, Japan}
\altaffiltext{16}{Department of Physics, Osaka City University, Osaka 558-858, Japan}

\email{\myemail}

\begin{abstract}
We searched for TeV gamma-rays from the remnant of SN 1987A
around 5400 days after the supernova.  
The observations were carried out in 2001, from November 16 to December 11, 
using the CANGAROO-II Imaging Atmospheric Cherenkov Telescope.
In total, 708 minutes of ON- and 1019 minutes of OFF-source data were obtained
under good conditions.
The detection threshold was estimated to be 1~TeV, due to the
mean zenith angle of 39$^\circ$.
The upper limits for the gamma-ray flux were obtained and compared with
the previous observations and theoretical models. 
The observations
indicate that the gamma-ray luminosity is lower than $1\times 10^{37}$
erg~s$^{-1}$ at $\sim 10$ TeV.
\end{abstract}


\section{Introduction} The explosion of SN 1987A on February 23, 1987,
in the Large Magellanic Cloud, was first detected as a short neutrino
burst \citep{koshiba87, hirata87, bionta87}. It was subsequently 
detected at almost all
wavelengths of the electromagnetic spectrum (see, e.g.,
\cite{chevalier92}, \cite{McCray93} and references therein).
After a weakening of
the emission, in accordance with the standard lightcurve for a core collapse
supernovae of Type II,  at present it shows a continuous increasing
brightness in radio \citep{manchester02} and X-ray bands \citep{park02}.

Although observational efforts in the high-energy gamma-ray region were
intensively carried out for several years
\citep{raubenheimer88, ciampa88, bond88a, bond88b, bond89, allen93a,
allen93b, stekelenborg93, yoshii96}, no positive signals were 
obtained, with the possible exception of
a 2-day TeV gamma-ray burst \citep{bond88b}. 
No observations since 1991 have been reported, despite the fact
that models predict an increasing flux of high energy gamma-rays
as the SN shock wave expands.

It is now 15 years since SN 1987A.
Even an upper limit for the current period of the supernova remnant
evolution would be very important to constrain models for gamma-ray
emission.
The technology used to detect very high-energy gamma-rays has improved
significantly over this 15 year period, particularly with the
development of Imaging Atmospheric
Cherenkov Telescopes (IACTs).
IACTs detect optical Cherenkov photons produced by electrons in 
cascades initiated by the interaction of gamma-rays at sub-TeV energies
in the Earth's upper atmosphere. The Cherenkov photons are strongly
beamed in the direction of the incident gamma-ray.
The CANGAROO-II telescope was, at the time of these observations,
the only one located in the southern hemisphere.

% Both with historical and scientific meaning, we should observe it.
% Even an upper limit should contribute to the physics.
% Time scale is $O$(10y) which in its log scale is important
% to the history of the universe.

The CANGAROO-II telescope is located near Woomera, South Australia.
Technical details are presented elsewhere
\citep{tanimori03}, and its performance is described in
\cite{enomoto02b, okumura02, itoh02}. The 10\,m diameter telescope
has an effective area of 57 m$^2$. SN~1987A can be seen
at a zenith angle of 38$^\circ$ at its culmination. 
As a result, we can measure the TeV region with a significantly better
sensitivity than in previous measurements.
We report here on the results of observations of SN~1987A.

\section{Observations and Analysis}

The observations were carried out in 2001 over ten moonless nights 
between November 16 and to December 11.
In total, 1275 min.\ of ON- and 1301 min.\ of OFF-source data were
recorded.
We removed the cloudy periods from the data and selected 708 min.\ of
ON- and 1019 min.\ of OFF-source data. The procedures and further 
details of the analysis can be found in \cite{itoh03}.

\begin{figure}
\plotone{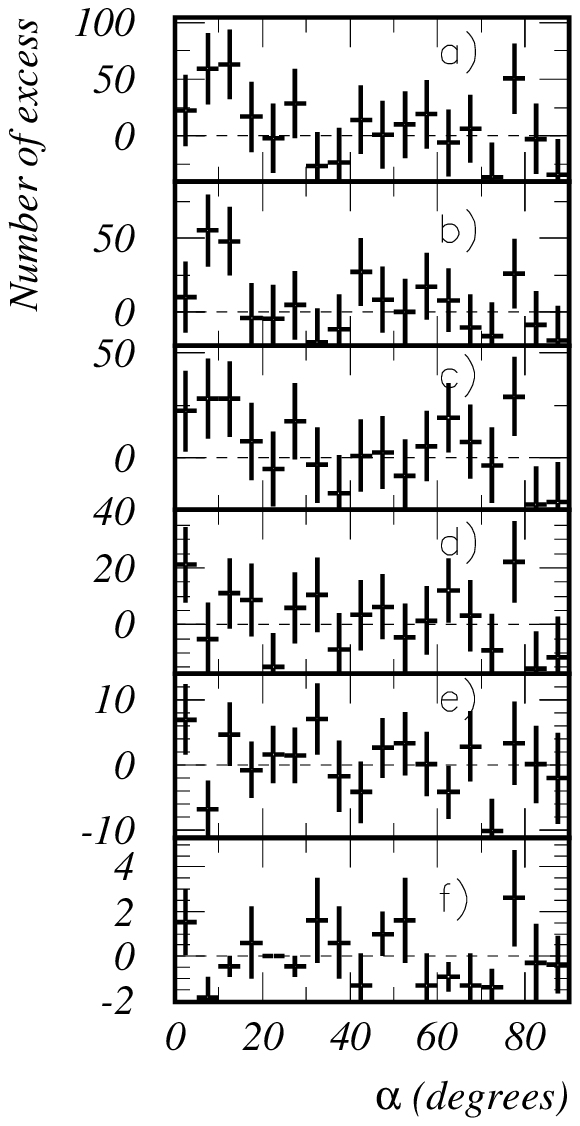}
\figcaption{Distributions of the image orientation angle ($\alpha$). 
These were obtained by
subtracting the normalized off-source data from the on-source data.
The ratio of
events in the higher $\alpha$ ($>$ 20$^\circ$)
regions for the on- and off-source data was used as
the normalization factor.
From top (a) to to bottom (f), six threshold values,
as shown in Table \ref{table1}, were applied to the analysis.\label{fig1}}

\end{figure}

Briefly, in the analysis, ``cleaning'' cuts   were applied to the
pixelized-camera images  (each pixel being 0.115$^\circ$ square), 
requiring that each pixel have greater than
$\sim$3.3~photoelectrons,  that the Cherenkov photons arrived within
$\pm$40~nsec, and that a cluster of at least five adjacent triggered 
pixels was contained in each event. 
After these pre-selections, we carried out a shower image
analysis using the standard set of image parameters, $distance$, $length$,
$width$,  and $\alpha$ \citep{hil85}, combining the 
$length$ and  $width$ (after
an initial $distance$ cut) to assign the likelihoods  to each event
\citep{enomoto02a}.  The  likelihoods for both a gamma-ray origin and
a cosmic-ray  proton origin were calculated. The cut that was used to reject
background events was based on the ratio of  these two likelihoods.
After these  cuts, the image orientation angles ($\alpha$) were plotted. 
A gamma-ray signal would appear as an excess at  low $\alpha$ after
the normalized OFF-source $\alpha$ distribution  is subtracted from the
ON-source  distribution. As shown in Figs.~\ref{fig1} a)--f), no
statistically significant excess of events 
%greater than a certain level (4--5$\sigma$) 
with  $\alpha < 15^\circ$ was observed. 
From top (a) to bottom (f) in Fig.~\ref{fig1}, six different
thresholds, which are shown in Table \ref{table1}, were applied to the
analysis. Our Monte Carlo simulations predicted that 73\% of the events
from a point source  would have $\alpha < 15^\circ$
at these zenith angles.

\section{Upper limits on the gamma-ray flux}

The upper limits to the emission at each energy
was obtained by adding the statistical and systematic errors
to any excess events in the relevant plot in Fig.~\ref{fig1}. 
The total error was the square root of the quadratic sum of both errors.
These errors were doubled to obtain $2\sigma$ Upper Limits (ULs).
In the case of a negative excess, 
only the errors were used to determine the upper limit.
%we forced it to be a small positive number.

The derivation of the integral flux depends on the unknown energy
spectrum of the incident gamma-rays.
%When deriving the integral flux, we suffered from the dependence
%of the energy spectrum of the incident gamma-rays.
We therefore tried several power-law energy spectra ($E^{-\gamma}$)
in Monte Carlo
simulations in order to determine the corresponding effective area of the
observations. Three spectra, with differential flux power-law indices of
$\gamma=2.0$, $2.5$, and $3.0$, were tested.
In all cases, the energy ranges of the generated gamma-rays
were 0.15--20~TeV.
We obtained integral flux upper limits under various assumptions,
as shown in Table \ref{table1}.
Although the threshold energies varied as expected 
with initial power-law indices,
the spectral responses roughly agreed with each other.
We therefore adopted $\gamma=2.0$, plotted
in Fig.~\ref{fig2} by the dotted line with arrows, together with the
previous measurements and model predictions.

\section{Discussion}

\begin{figure}
\plotone{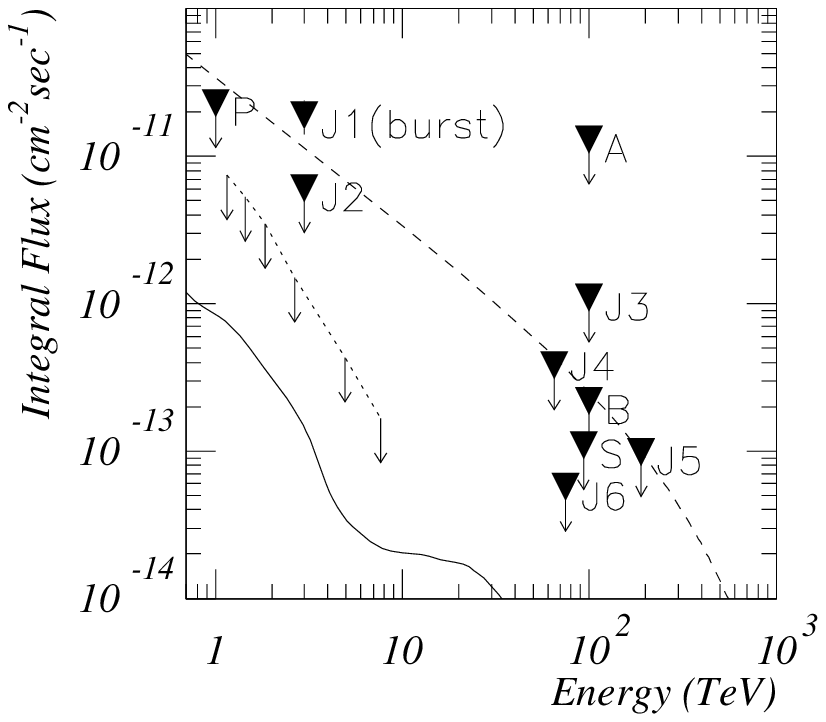}
\figcaption{Upper limits on the flux of gamma-rays. The dotted line
is that obtained by this experiment. P: \cite{raubenheimer88}; J1--6:
\cite{bond88a, bond88b, bond89, allen93a, allen93b}; A: \cite{ciampa88};
S: \cite{stekelenborg93}; B: \cite{yoshii96}. The solid and dashed lines
are theoretical predictions of the flux by \cite{berezhko00, gaisser89},
respectively.
\label{fig2}
}
\end{figure}

In Fig.~\ref{fig2} we compile the upper limits on the flux of gamma rays
from SN 1987A of this observation (dotted line with arrows) and those 
reported by others at different times since the explosion. Theoretical
predictions by \cite{berezhko00} (solid line), which correspond to a time
of $\sim$5000 days, and by \cite{gaisser89} (dashed line), which is almost
constant in time, are also shown.

The upper limits of this observation are significantly better than those of
previous observations. In particular, at 3~TeV it is a factor of 20
lower than that of \cite{bond88b}. % , who claimed a burst.
At 1~TeV, the upper limit is tightened by a factor of 3, 
and at the highest point
(several TeV) it is improved by a factor of 50.
Previous measurements calculated typical luminosity upper limits of 
several times 10$^{38}$ erg\,s$^{-1}$, using a distance of 
$\sim$50~kpc.
This observation indicates that the TeV gamma-ray luminosity is 
lower than $1\times 10^{37}$~erg\,s$^{-1}$ at $\sim$10~TeV, which is now 
of a similar order to
those in bright high-energy astronomical objects at 
various wavelengths.

The predictions concerning the  emitted high-energy gamma-rays from
this source have been extensively discussed  \citep{honda87,
nakamura87, yamada88, berezinsky89, gaisser89, schlickeiser91,
berezhko00}. High-energy  photons can be produced in collisions of
accelerated particles with the ambient medium. There are several
processes which could accelerate particles in young supernova remnants 
\citep[see, e.g.,][for a review]{dogiel89}.

\cite{gaisser89} discuss the acceleration of
particles at the pulsar wind shock. However, an analysis of the
2.14~ms pulsar candidate in the remnant of SN 1987A
\citep{middleditch00} suggests that the magnetic field strength at the
surface of a neutron star has an upper limit of $\sim 10^{10}$~G
\citep{nagataki01}, which is about two orders less than typical values
and that assumed by \cite{gaisser89}.

The lightcurve of soft X-rays, which are expected from the interaction of
the supernova shock with the matter, can be well fitted with a $t^2$
relation \citep{aschenbach02}. The recent X-ray data points tend to
exceed the $t^2$ best fit \citep{aschenbach02, park02}. One can expect
the similar behaviour of the TeV gamma-ray flux from collisions 
of accelerated cosmic rays with the ambient matter.

\begin{figure}
\plotone{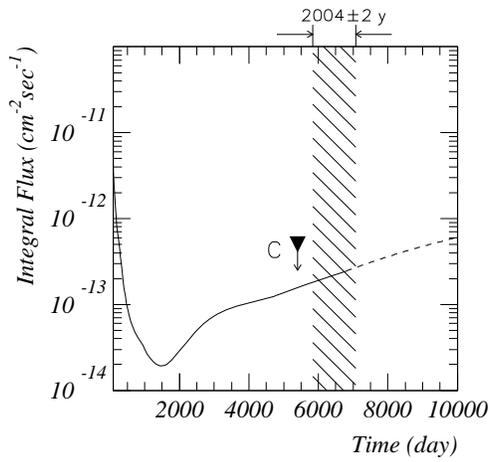}
\figcaption{Flux of gamma-rays with energy $E>3$~TeV vs. time since the
SN~1987A explosion. The current work upper limit (C) is shown.  
The solid curve was extracted from \cite{berezhko00}. 
The dashed curve is an extrapolation to the solid one.
The time when the shock is expected to encounter the inner
optical ring is the hatched region. 
\label{fig3}
}
\end{figure}

Fig.~\ref{fig3} shows the dependence of the gamma-ray 
flux with an energy $> 3$~TeV on time since the explosion. 
The solid line is extracted
from the results of numerical calculations by \cite{berezhko00}. The dashed
curve is an extrapolation to that curve under the assumption that
$F_{\gamma} \propto t^2$, which is a reasonable lower limit of the expected
flux in the future. One can see that the our 
upper limit is just a factor of 3 above the theoretical prediction for the
current epoch.

The presence and growing amount of synchrotron radio emission
unambiguously testify to the presence of high-energy electrons accelerated
by the forward-moving ejecta-driven shock. The radio spectrum has a
power-law index of 0.88 \citep{manchester02}, which is much softer than
the value of 0.5 from linear diffusive shock acceleration. This can be
explained with an essential modification of the shock wave due to the very
effectively accelerated nucleonic cosmic ray pressure
\citep{berezhko00}. Also, radio measurements show that the shock velocity
has dropped from
the initial value of 10000--35000~km\,s$^{-1}$ to $\sim$3000~km\,s$^{-1}$ 
\citep{gaensler97},
which is consistent with the shock having encountered a denser shocked
component of the progenitor's stellar wind with a number density of $\sim
100$~cm$^{-3}$  \citep{chevalier95}. It is thus reasonable to expect 
a considerable flux of TeV gamma-rays from the decay of $\pi^0$ mesons
produced in collisions of the cosmic ray nucleonic component with the ambient
matter nuclei \citep{drury94, naito94}, which is expected to increase
approximately by a factor of 2 between 2000 and 2006 \citep{berezhko00}. 

At the current rate of
expansion, the shock will encounter the much denser inner
optical ring in the year 2004$\pm$2 \citep{manchester02}. Thus, one can
also expect a dramatic increase of the TeV gamma-ray flux, which
could well exceed the current upper limits. The future
detection of TeV gamma-ray emission, 
%which will be higher than the current or
%future lower upper limits, 
will unambiguously prove the idea that
the main part of nucleonic cosmic rays are indeed accelerated at the supernova
remnant shock waves by a diffusive acceleration process. 

%30 Dor C: The total non-thermal luminosity is $7\times 10^{35}$ erg/s,

The next generation of southern
hemisphere IACTs, CANGAROO-III and H.E.S.S., will have improved sensitivities
and reduced energy thresholds. Considering the 
present theoretical estimations and recent radio and X-ray
observations, deep observations, with a total ON-source exposure of 
$\sim$100 hours, will have a good chance of detecting a signal.
A more detailed theory of high-energy gamma-ray production in the SN 1987A
environment is now needed.
Regular observations over the next decade are also highly desirable.

\acknowledgments

This work was supported by a Grant-in-Aid for Scientific Research by
the Ministry of Education, Culture, Science, Sports and Technology of
Japan and Australian Research Council. The receipt of JSPS Research
Fellowships is also acknowledged.

\clearpage

\clearpage

%\begin{deluxetable}{rrrrrrr}
\begin{deluxetable}{ccccccc}
\tablecolumns{7}
\tablewidth{0pc}
\tablecaption{Integral flux upper limit ($2\sigma$). \label{table1}}
\tablehead{
\colhead{} & \colhead{$\gamma$ = 2.0} & \colhead{}
           & \colhead{$\gamma$ = 2.5} & \colhead{}
           & \colhead{$\gamma$ = 3.0} & \colhead{} \\
\cline{2-2} \cline{4-4} \cline{6-6}\\
\colhead{bin} & \colhead{$E_{threshold}$} & \colhead{$2\sigma$-UL}
              & \colhead{$E_{threshold}$} & \colhead{$2\sigma$-UL}
              & \colhead{$E_{threshold}$} & \colhead{$2\sigma$-UL}\\
\colhead{number} & \colhead{(TeV)} & \colhead{(cm$^{-2}$s$^{-1}$)}
                 & \colhead{(TeV)} & \colhead{(cm$^{-2}$s$^{-1}$)}
                 & \colhead{(TeV)} & \colhead{(cm$^{-2}$s$^{-1}$)}}
\startdata
1 & 1.2 & 7.5$\times 10^{-12}$ 
  & 1.0 & 1.0$\times 10^{-11}$ 
  & 0.9 & 1.3$\times 10^{-11}$ \\
2 & 1.5 & 5.3$\times 10^{-12}$ 
  & 1.4 & 5.8$\times 10^{-12}$ 
  & 1.3 & 6.5$\times 10^{-12}$ \\
3 & 1.9 & 3.5$\times 10^{-12}$ 
  & 1.7 & 3.9$\times 10^{-12}$ 
  & 1.7 & 3.7$\times 10^{-12}$ \\
4 & 2.7 & 1.5$\times 10^{-12}$ 
  & 2.2 & 1.8$\times 10^{-12}$ 
  & 2.4 & 1.4$\times 10^{-12}$ \\
5 & 5.0 & 4.3$\times 10^{-13}$ 
  & 4.0 & 4.8$\times 10^{-13}$ 
  & 3.5 & 5.0$\times 10^{-13}$ \\
6 & 7.7 & 1.7$\times 10^{-13}$ 
  & 8.0 & 1.3$\times 10^{-13}$ 
  & 8.2 & 8.5$\times 10^{-14}$ 
\enddata
\end{deluxetable}

\end{document}